\documentclass[conference]{IEEEtran}
\usepackage[T1]{fontenc}
\usepackage{txfonts}
\usepackage{graphicx}

\newcommand{\ket}[1]{|#1\rangle}
\newcommand{\bra}[1]{\langle #1 |}
\newtheorem{remark}{Remark}

\newcommand{\Tr}{\mathrm{Tr}}

\begin{document}
\title{Improved Asymptotic Key Rate of the B92 Protocol}
\author{\IEEEauthorblockN{Ryutaroh Matsumoto\IEEEauthorrefmark{1}}
\IEEEauthorblockA{\IEEEauthorblockA{\IEEEauthorrefmark{1}Department of Communications and Integrated Systems, Tokyo Institute of Technology, 152-8550 Japan}}}
\date{January 22, 2013}
\maketitle

\begin{abstract}
We analyze the asymptotic key rate of the
single photon B92 protocol by using 
Renner's security analysis given in 2005.
The new analysis shows that the B92 protocol
can securely generate key at
6.5\% depolarizing rate, while the previous analyses 
cannot guarantee the secure key generation at 4.2\% depolarizing rate.
\end{abstract}

\section{Introduction}
The B92 quantum-key-distribution (QKD) protocol
\cite{bennett92qkd} has remained less popular than the
famous BB84 protocol \cite{bennett84},
while both protocols provide the unconditional security.
One plausible reason for the unpopularity is that
the B92 is weaker to the channel noise than the BB84.
Specifically, the BB84 with the standard one-way
information reconciliation can generate secure key
over the depolarizing channel at  depolarizing rate
16.5\%, while
the previous  security analyses of the B92
cannot guarantee the secure key generation at
depolarizing rate 3.5\% \cite{tamaki03},
3.7\% \cite{genericqkd} or 4.2\% \cite{renner05}.

The conventional security analyses of
the B92 \cite{genericqkd,renner05,tamaki03}
involved many inequalities, and the tightness of those
inequalities was not explicitly discussed.
We cannot exclude the possibility that the B92 protocol
can securely generate key at depolarizing rates over 4.2\%.
On the other hand,
the asymptotic secure key rate in \cite{renner05,rennerphd} is expressed as
the minimum of conditional quantum entropy over a certain set
of bipartite quantum states.
By using the convex optimization technique,
we can completely remove careful manipulation of
many inequalities, which could underestimate the
secure key rate.
In this paper
we reformulate the asymptotic secure key rate formula as
a convex optimization problem,
and compute the rate without any manipulation of inequalities
directly by a numerical optimization procedure.
As a result, we show that the B92 protocol \cite{bennett92qkd}
without
noisy preprocessing \cite{renner05} can securely
generate key at 6.5\% depolarizing rate.

\section{New Security Analysis of the B92 Protocol}
In this section, we present
a new formula for the asymptotic key rate of the B92 protocol,
based on Renner's security argument \cite{rennerphd}.
Firstly, we fix notations.
Let $\{\ket{0}$, $\ket{1}\}$ be some fixed orthonormal
basis of a qubit.
In the B92 protocol \cite{bennett92qkd},
Alice sends the quantum state
\begin{equation}
\ket{\varphi_j} = \beta\ket{0}+(-1)^j\alpha\ket{1}, \label{eq:alpha}
\end{equation}
for $j=0,1$, where $\beta = \sqrt{1-\alpha^2}$,
and $0 < \alpha< 1/\sqrt{2}$.
For convenience of presentation, we also define
\[
\ket{\bar{\varphi}_j} =\alpha\ket{0} - (-1)^j \beta\ket{1}.
\]
We can see that $\{\ket{\varphi_j}$, $\ket{\bar{\varphi}_j}\}$ forms
an orthonormal basis of a qubit.

On the other hand,
we can express a qubit channel as follows.
Define the three Pauli matrices $\sigma_x$, $\sigma_y$,
and $\sigma_z$ as usual.
Then a qubit density matrix $\rho$ can be expressed
as \cite{chuangnielsen}
\[
\rho = \frac{1}{2}\left(
I + x \sigma_x + y \sigma_y + z \sigma_z\right),
\]
where $x,y,z\in \mathbf{R}$ and
$x^2 + y^2 + z^2 \leq 1$.
The vector $(x,y,z)$ is called a Bloch vector.
The qubit channel $\mathcal{E}_B$ from Alice
to Bob can be expressed as a map between
Bloch vectors by
\begin{equation}
\left(\begin{array}{c}
z\\
x\\
y
\end{array}\right)
\mapsto
R \left(\begin{array}{c}
z\\
x\\
y
\end{array}\right) + \vec{t},\label{eq:channelparameter}
\end{equation}
where
\begin{equation}
R = \left(
\begin{array}{ccc}
R_{zz}&R_{zx}&R_{zy}\\
R_{xz}&R_{xx}&R_{xy}\\
R_{yz}&R_{yx}&R_{yy}
\end{array}
\right), \quad
\vec{t} = 
\left(\begin{array}{c}
t_z\\
t_x\\
t_y
\end{array}\right). \label{eq:parameters}
\end{equation}

Define
\[
\ket{\Psi}
= \frac{\ket{0}_A\ket{\varphi_0}_B + \ket{1}_A\ket{\varphi_1}_B}{\sqrt{2}}.
\]
As in \cite{tamaki03},
we also define the four POVM 
\begin{eqnarray}
F_0 &=& \ket{\bar{\varphi}_1}\bra{\bar{\varphi}_1}/2,\label{povm1}\\
F_1 &=& \ket{\bar{\varphi}_0}\bra{\bar{\varphi}_0}/2,\\
F_{\bar{0}} &=& \ket{{\varphi}_1}\bra{{\varphi}_1}/2,\\
F_{\bar{1}} &=& \ket{{\varphi}_0}\bra{{\varphi}_0}/2.\label{povm4}
\end{eqnarray}

After passing the quantum channel $\mathcal{E}_B$
from Alice to Bob,
$\ket{\Psi}\bra{\Psi}$ becomes
\begin{equation}
\rho_{1,AB} = (I \otimes \mathcal{E}_B) \ket{\Psi}\bra{\Psi}.\label{eq:afterchannel}
\end{equation}
In a quantum key distribution protocol,
the state change $\mathcal{E}_B$ is caused by Eve's cloning of the
transmitted qubits to her quantum memory. The content of Eve's quantum
memory is mathematically described by the purification
$\ket{\Phi_{1,ABE}}$ of $\rho_{1,AB}$.
Let $\rho_{1,ABE} = \ket{\Phi_{1,ABE}}\bra{\Phi_{1,ABE}}$.

In addition to Eve's quantum memory,
she also knows the content of public communication over the
classical public channel between Alice and Bob.
For each transmitted qubit from Alice to Bob,
the public communication consists of $1$-bit information
indicating whether Bob discards his received qubit or not.
We also have to take it into account. We shall represent the
public communication by a classical random variable $P$
that becomes $1$ if Bob discards his qubit and $0$ otherwise.
So, $P=0$ when Bob's measurement outcome is $F_0$ or $F_1$,
and $P=1$ when Bob's measurement outcome is $F_{\bar{0}}$ or $F_{\bar{1}}$.

On the other hand, in the B92 protocol,
Bob performs the measurement specified by Eqs.\ (\ref{povm1})--(\ref{povm4}).
Alice and Bob keep their a qubit if and only if its measurement outcome
is $F_0$ or $F_1$. Otherwise it is discarded and is not used for
generation of secret key.
This is mathematically equivalent to set Alice's bit to $0$ if
the measurement outcomes is $F_{\bar{0}}$ or $F_{\bar{1}}$.
Therefore, from Eve's perspective on Alice's classical bit,
the joint state between Alice and Bob after the selection by measurement
outcomes is equivalent to
\begin{eqnarray*}
\rho_{2,ABEP} &=& (I_{A}\otimes \sqrt{F_0} \otimes I_E\rho_{1,ABE} I_{A}\otimes \sqrt{F_0}\otimes I_E\\ && + I_{A}\otimes \sqrt{F_1}\otimes I_E\rho_{1,ABE} I_A\otimes \sqrt{F_1}\otimes I_E)\otimes \ket{0}_P\bra{0}_P \\
&& +
\ket{0}_A\bra{0}_A\otimes (\sqrt{F_{\bar{0}}}\otimes I_E\Tr_A[\rho_{1,ABE}]\sqrt{F_{\bar{0}}}\otimes I_E\\ &&
 + \sqrt{F_{\bar{1}}}\otimes I_E\Tr_A[\rho_{1,ABE}]\sqrt{F_{\bar{1}}}\otimes I_E)\otimes \ket{1}_P\bra{1}_P. \label{eq:afterselection}
\end{eqnarray*}
Observe that the state change from $\rho_{1,ABE}$ to
$\rho_{2,ABEP}$ is a trace-preserving completely positive map.

\begin{remark}
Alternatively,
by using the more usual approach to model a quantum state after 
selective measurement,
one can also regard the quantum state after having Bob's
measurement outcome $F_0$ or $F_1$ as
\begin{eqnarray*}
&&\frac{1}{(F_0 + F_1) \mathrm{Tr}_{A}[\rho_{1,AB}]}
(I_{A}\otimes \sqrt{F_0} \otimes I_E\rho_{1,ABE} I_{A}\otimes \sqrt{F_0}\otimes I_E\\ && + I_{A}\otimes \sqrt{F_1}\otimes I_E\rho_{1,ABE} I_A\otimes \sqrt{F_1}\otimes I_E)\otimes \ket{0}_P\bra{0}_P.
\end{eqnarray*}
The motivation behind using our alternative formulation (\ref{eq:afterselection})
is to prove later the convexity of the quantum conditional entropy
(\ref{eq:sxe})
in terms of the parameters given in Eq.\ (\ref{eq:parameters}),
so that we can use the convex optimization technique to
find the minimum value of Eq.\ (\ref{eq:sxe}).
\end{remark}

In order to calculate the key rate, we need to consider
Eve's ambiguity on Alice's classical bit \cite{renner05,rennerphd}
defined as follows. 
Let
\[
\rho_{2,XEP} = \sum_{j=0,1} \ket{j}_A\bra{j}_A\otimes I_{EP} \mathrm{Tr}_B[\rho_{2,ABEP}] \ket{j}_A\bra{j}_A\otimes I_{EP}.
\]
Eve's ambiguity on Alice's classical bit $S(X|EP)$ is defined as
\begin{equation}
S(X|EP) = S(\rho_{2,XEP}) - S(\rho_{2,EP}),\label{eq:sxe}
\end{equation}
where $\rho_{2,EP} = \Tr_A[\rho_{2,XEP}]$, and $S(\cdot)$ denotes
the von Neumann entropy.

In order to calculate the amount of public communication required for
information reconciliation, we define the joint random variables
$(X',Y')$ as
\begin{eqnarray}
X'&=& j\textrm{ if the transmitted qubit is }\ket{\varphi_j}, \nonumber\\
Y'&=& k \textrm{ if the measurement outcome is }F_k,\label{eq:y}
\end{eqnarray}
under the condition that the measurement outcome is
either $F_0$ or $F_1$.
Observe the difference between $X$ and $X'$.
$X'$ is not defined  but $X$ is defined to be $0$
when Bob's measurement outcome is either 
$F_{\bar{0}}$ or $F_{\bar{1}}$.

We shall show the asymptotic key rate per single transmitted
qubit that is neither announced for the channel estimation nor
discarded due to the measurement outcome being $F_{\bar{0}}$ or $F_{\bar{1}}$.
Note that Eq.\ (\ref{eq:sxe}) is Eve's ambiguity per a qubit
that is not announced for the channel estimation but \emph{can be
discarded}.
The probability of the measurement outcome being
$F_0$ or $F_1$ is
\[
\Tr[\rho_{1,AB} (I_A \otimes (F_0 + F_1))].
\]
So we can see that Eve's ambiguity  per single transmitted
qubit that is neither announced for the channel estimation nor
discarded is
\[
\frac{S(X|EP)}{\Tr[\rho_{1,AB} (I \otimes (F_0 + F_1))]}.
\]
By \cite{renner05,rennerphd} the asymptotic key rate is
\begin{equation}
\frac{S(X|EP)}{\Tr[\rho_{1,AB} (I \otimes (F_0 + F_1))]} - H(X'|Y').\label{eq:keyrate}
\end{equation}

Note that the above formula assumes that
Alice and Bob knows the channel between them.
In the BB92 protocol,
we cannot estimate all the parameters of the channel.
We can only estimate part of them.
In Eq.\ (\ref{eq:keyrate}) we can asymptotically determine the
true values of $\Tr[\rho_{1,AB} (I \otimes (F_0 + F_1))]$ and
$H(X'|Y')$.
On the other hand we cannot know the true value of $S(X|EP)$.
Therefore, we need to calculate
the minimum value (i.e.\ the worst-case) of $S(X|EP)$
over all the possible quantum channel $\mathcal{E}_B$ between
them.

One can compute the minimum of $S(X|EP)$ as follows.
Observe first that $S(X|EP)$ is a function of the channel 
parameters Eq.\ (\ref{eq:parameters}) of $\mathcal{E}_B$.
By the almost same argument as \cite[Remark 11]{watanabe08}
one sees that $S(X|EP)$ is a convex function of the channel 
parameters Eq.\ (\ref{eq:parameters}).
Moreover,
we see that the minimum of $S(X|EP)$ is attained 
when $R_{xy} = R_{yx} = R_{yz} = R_{zy} = t_y = 0$
by the almost same argument as \cite[Proposition 1]{watanabe08}.
Therefore, one can compute the minimization of $S(X|EP)$ by
the convex optimization \cite{boyd04}.



\section{Numerical Result}
We consider the depolarizing channel $\mathcal{E}_q$
with
depolarizing rate $q$.
The definition of $q$ follows \cite{tamaki03}.
For a qubit density matrix $\rho$,
we have $\mathcal{E}_q(\rho) = (1-q) \rho + (q/2) I_{2 \times 2}$.
With such a channel $\mathcal{E}_q$,
$R$ and $\vec{t}$ in Eq.\ (\ref{eq:channelparameter})
are given by
\[
R = \left(
\begin{array}{ccc}
1-4q/3&0&0\\
0&1-4q/3&\\
0&0&1-4q/3
\end{array}
\right), \quad
\vec{t} =  \vec{0}.
\]
Define
\[
\rho_{1,AB,q} = (I \otimes \mathcal{E}_q) \ket{\Psi}\bra{\Psi}.
\]
Over $\mathcal{E}_q$ with infinitely many qubits,
the asymptotic key rate is given by
\begin{equation}
\frac{\min S(X|EP)}{\Tr[\rho_{1,AB} (I \otimes (F_0 + F_1))]} - H(X'|Y'),
\label{eq:min}
\end{equation}
where the minimum is taken over the set of parameters in Eq.\ (\ref{eq:parameters})
such that
\begin{eqnarray}
&&\Tr[(\ket{0}\bra{0}\otimes F_0+\ket{1}\bra{1}\otimes F_1)\rho_{1,AB}] \nonumber\\
&=& 
\Tr[(\ket{0}\bra{0}\otimes F_0+\ket{1}\bra{1}\otimes F_1)\rho_{1,AB,q}],\label{eq11}\\
&&\Tr[(\ket{1}\bra{1}\otimes F_0+\ket{0}\bra{0}\otimes F_1)\rho_{1,AB}] \nonumber\\
&=& 
\Tr[(\ket{1}\bra{1}\otimes F_0+\ket{0}\bra{0}\otimes F_1)\rho_{1,AB,q}].\label{eq12}
\end{eqnarray}
We also required that parameters in Eq.\ (\ref{eq:parameters})
represent a completely positive map.
We stress that we do not restrict the range of minimization to
the depolarizing or the Pauli channels. 
The minimization is carried out over the set of all the qubit channels
with (\ref{eq11}) and (\ref{eq12}).

The FindMinimum function
in Mathematica 8.04 was used for the minimization.
The program source code and the computation results are
included in this eprint.

\begin{figure}[t!]
\includegraphics[width=\linewidth]{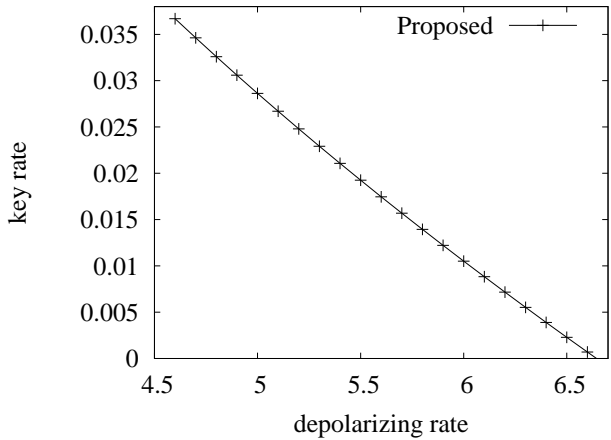}
\caption{Asymptotic Key Rate: The conventional methods
\cite{genericqkd,renner05,tamaki03}
cannot generate key at
depolarizing rate above 4.2\% and they are not plotted.}\label{fig1}
\end{figure}

We only considered $\alpha=0.39$ and did not optimized
the value of $\alpha$ in Eq.\ (\ref{eq:alpha}).
The key rate is plotted in Fig.\ \ref{fig1}.
The convex optimization did not converge in $10^5$ iterations
when the depolarizing rate $\leq 4.5\%$. The key rate is plotted from
depolarizing rate $ \geq
4.6\%$.


\section{Conclusion}
In this paper, we reformulated the secure key rate formula
of the B92 protocol as a convex optimization.
We have not resorted to skillful manipulation of inequalities,
and the secure key rate is computed simply by a numerical
optimization procedure. The result shows that the B92 protocol
can securely generate key at significantly higher depolarizing
rates than previous security analyzes.

\section*{Acknowledgment}
The author would like to thank
K. Azuma, G. Kato, K. Tamaki and T. Tsurumaru for helpful discussions.
This research is partly supported by NICT and JSPS.



\end{document}